\documentstyle[12pt,epsf]{article}

\catcode`\@=11
\def\numberbysection{\@addtoreset{equation}{section}
        \def\theequation{\thesection.\arabic{equation}}}

\newcommand{\be}{\begin{equation}}
\newcommand{\fey}[1]{#1\hspace{-.78em}/\hspace{-.07em}}
\newcommand{\ee}{\end{equation}}
\newcommand{\ba}{\begin{eqnarray}}
\newcommand{\ea}{\end{eqnarray}}

\newcommand{\nl}{\nonumber \\}

\def\ra{\rangle}
\def\la{\langle}

\def\a{\alpha}
\def\b{\beta}

\def\G{\Gamma}
\def\D{\Delta}
\def\d{\delta}

\def\z{\zeta}

\def\l{\lambda}

\def\m{\mu}
\def\n{\nu}

\def\p{\pi}

\def\r{\rho}
\def\s{\sigma}

\def\f{\phi}

\numberbysection

\begin{document}
\begin{titlepage}
\begin{center}
\hfill  \quad DFF 354/05/2000 \\
\hfill  \quad hep-th/0005115 \\

\vspace{1.5cm}

{\LARGE On the Trace Anomaly }

\vspace{.3cm}

{\LARGE as a Measure of Degrees of Freedom}

\vspace{1.cm}

\qquad Andrea CAPPELLI \quad {\it and} \quad
Giuseppe D'APPOLLONIO \\
\vskip 0.1in
{\em I.N.F.N. and Dipartimento di Fisica}\\
{\em  Largo E. Fermi 2, I-50125 Firenze, Italy}
\end{center}

\vspace{.5cm}

\begin{abstract}
Recent conjectures of the $c$-theorem in four and higher dimensions
have suggested that the coefficient of the Euler characteristic
in the trace anomaly could measure the degrees of
freedom in field theory and decrease along the renormalization-group flow.
We compute this quantity for free massless scalar, fermion and 
antisymmetric tensor fields in any dimension, 
and analyse its dependence on spin and space-time dimension.
In the limit of large number of dimensions, where the theories 
become semiclassical, we find that this quantity does not approach
the classical number of field components, but is enhanced for
spinful particles.
This seemingly strange behaviour is found to be
consistent with known renormalization-group patterns and
a specific $c$-theorem conjecture.
\end{abstract}

\vfill
\hfill May 2000 
\end{titlepage}
\pagenumbering{arabic}


\section{Introduction}

The $c$-theorem establishes the irreversibility of the 
renormalization-group flow in two dimensions \cite{cth}; it shows
that there exists a positive function of the coupling constants
which monotonically decreases along the flows
and is stationary at the fixed points. 
The existence of an analogous theorem in four and higher dimensions
\cite{cardy}\cite{cfl}
has been extensively discussed in the recent literature:
several non-trivial flows have been checked
in four-dimensional supersymmetric gauge theories (most notably
in the ``conformal window'') \cite{afgj}, and {\it via}
the AdS/CFT correspondence \cite{fgpw}; moreover, two proofs of the theorem
have been presented in the Refs.\cite{fl}\cite{ath}.

Most of the examples and conjectures identify the $c$-function
at the fixed points with the coefficient of the Euler density $G$ in the
trace anomaly (such that $\chi=\int G$ is the Euler characteristic).
This coefficient, called $a$, 
has definite sign for all unitary theories \cite{ol},
and verify the inequality $a_{UV} > a_{IR}$
for all known examples of flows connecting pairs of UV and IR fixed points.

According to the analysis of Ref.\cite{ds},  
the trace anomaly in any even dimension $d=2k$ contains three parts:
the topological invariant term, given by the Euler density;
the Weyl invariant terms, made by polynomials of the Weyl tensor
and its derivatives, and finally the variations of
local counterterms which can be neglected.
On conformally-flat geometries, like the $d$-dimensional sphere $S^{2k}$,
the Weyl tensor vanishes and the trace anomaly is completely
expressed in terms of the Euler characteristic.

In this paper we calculate the trace anomaly on the sphere $S^{2k}$
for the free massless scalar, Dirac spinor and antisymmetric tensor 
fields in any dimension and obtain the corresponding values of $a$.  
We use the standard zeta-function regularization  of the
partition function \cite{hawk}.

Next, we analyse the behaviour of $a$ upon varying the spin and the
dimension and test its interpretation as a measure of 
degrees of freedom in field theory.
In the large-$d$ limit, we find that all anomalies, once properly 
normalized, go to zero; this is consistent with the expected
semiclassical behaviour. 
However, the ratios $a(\s)/a(0)$ of the spinful to the scalar values, which
have the meaning of counting functions, do not approach the
classical number of fields components:
each component of the antisymmetric
tensor field weights of order $O(d^3)$ with respect to
the scalar field, and each fermion component weights $O(d)$.
Therefore, the anomaly $a$ yields a ``quantum'' measure of degrees of
freedom which is rather far from the classical intuition, with the
antisymmetric tensor dominating the counting for any $d \ge 4$.

At first sight, we could doubt of this 
interpretation as a measure of degrees of freedom;
however, a closer inspection shows that this is consistent 
with known facts and renormalization-group flows.
The same enhancement is found for the 
gravitational chiral anomaly \cite{lagw}
and for another term in the trace anomaly, whose
coefficient $c$ also normalizes the stress-tensor two-point 
function\footnote{
$c$ is always positive, but is not always decreasing
 \cite{cfl}\cite{afgj}.}
\cite{cfl}.
Furthermore, the observed dominance of the antisymmetric tensor field 
is crucial for the existence of theories with $c\propto a$
in any dimension \cite{c-a}; these theories naturally arise in 
the AdS/CFT correspondence \cite{hs}\cite{agpz} and
have been conjectured to satisfy a version of $c$-theorem
with all its two-dimensional features \cite{c-a}.


\section{Trace anomaly of the scalar and Dirac fermion fields}

The stress tensor $T_{\m \n}$ of a $d$-dimensional field theory is 
defined by the variation of the partition function with respect to the
the background metric $g_{\m \n}\ \ $:
\be
\d \log Z[g_{\m \n}] =   
\frac{1}{2V_{d-1}}\int dx^d\ \sqrt{g} \ T_{\m \n} {\d g^{\m \n}}\ , 
\qquad\qquad V_{d-1} = \frac{2 \p^{d/2}}{\G(\frac{d}{2})}\ ,
\label{tdef}\ee
where the conventional normalization factor $V_{d-1}$ is the volume of 
the unit sphere $S^{d-1}$.
We shall compute the partition function of $d$-dimensional
free field theories on the sphere $S^{2k}$ ($d=2k$) and obtain the 
integrated trace anomaly by the scale variation 
$g_{\m \n} \to \exp(2\a )\ g_{\m \n}\ $ ($\a={\rm const.}$):
\be
\frac{d}{d \a} \log Z \left[ S^{2k} \right] \ = \ 
- \frac{1}{V_{2k-1}} \int_{S^{2k}} \la \Theta \ra \ =\ 
- { V_{2k} \ r^{2k} \over V_{2k-1}} \ \la \Theta \ra \ .
\label{thedef}\ee 
In this Equation, $\Theta$ is a short-hand notation for the trace 
$T_\m^\m$ and $r $ is the radius of the sphere.

The (Euclidean) action for bosonic free fields is written in terms of 
the covariant Laplacian $\D$, whose explicit form depends on the
spin of the field:
\be
S[g,\f] = \frac{1}{2} \int \sqrt{g}\ \f(x)\ \D\ \f(x) \ .
\label{Sdef}\ee
The partition function, i.e. the determinant of the Laplacian, can 
be obtained from the analytic continuation of the zeta-function
\cite{hawk} associated to the spectrum of $\D$ on $S^{2k}$, 
\be
\z_{\D}(s) = \sum_{n=0}^{\infty}\ \frac{\d_n}{\l^s_n} \ ;
\label{zetadef}\ee
this function is defined in terms of the eigenvalues $\l_n$ and
their degeneracies $\d_n$.

We consider theories which are invariant under conformal (Weyl) 
transformations $g_{\mu\nu}(x)\to \exp 2\s (x)\ g_{\mu\nu}(x)$ at the
classical level \cite{bd}\cite{cc}; the corresponding
Laplacian is conformally covariant,
$ \D \to \exp(-(\d +2)\s)\ \D\ \exp(\d \s) $,
with $\d$ the dimension of $\phi$.
In these theories, the scale dependence of the partition 
function (\ref{thedef}) 
is purely anomalous, i.e. is induced by the regularization of the
determinant and it is given by the zeta-function
analytically continued at $s=0$ \cite{hawk}:
\be
\frac{d}{d \a} \log Z \left[ S^{2k} \right] \ = \ 
\ \z_{\D}(0)\ \ =
- \frac{1}{V_{2k-1}} \int_{S^{2k}} \la \Theta \ra \ .
\label{zeta0}
\ee
In the case of fermionic fields, we consider the action (\ref{Sdef})
with $\D$ the square of the Dirac operator and obtain
the anomaly (\ref{zeta0}) in terms of $- \z_{\D}(0)$.


We now evaluate the trace anomaly for the scalar 
field\footnote{
This calculation and the following one for the Dirac fermion
are not new \cite{ct1}, but we 
present here rather compact results.},
whose conformal covariant Laplacian is \cite{bd}:
\be
\D_{0} = - \nabla^2 + \frac{k-1}{2(2k-1)}{\cal R} \ ,
\label{a0}\ee
with ${\cal R}$ the scalar curvature.
The eigenvalues and degeneracies of this operator on $S^{2k}$
are given by 
\ba
\l_{l} &=& (l+k)(l+k-1) \quad , \ \hspace{3cm} l = 0, 1, 2, ... \nonumber \\
\d_{l} &=& (2l+2k-1) \frac{(l+2k-2)!}{l!(2k-1)!}  \quad .
\label{eigen0}
\ea
The zeta-function of the scalar Laplacian $\z_{\D_0}(s)$ 
(\ref{zetadef},\ref{eigen0}) can be analytically continued
at $s=0$ by expressing it in terms of ordinary 
Riemann zeta-functions $\z (s)$. 
It is convenient to use the analytic continuation formula 
found in Ref.\cite{weis}:
\ba
\sum_{\s=n+1}^{\infty}
\frac{2 \s + 1}{\left[(\s - n)(\s + n + 1)\right]^{s}}
& = & 
\frac{1}{\G(s)} \sum_{m=0}^{\infty} 
\frac{\G(s\!+\!m\!\!-\!\!1)}{\G(m+1)}(2s+m-2)(2n+1)^m \nl
&& \times \left[\z(2s+m-1)-\sum_{l=1}^{2n+1}l^{-2s-m+1} \right] \ .
\label{weized}
\ea
We rewrite the zeta-function of the scalar Laplacian as follows
(for $k\ge 2$):
\be
\z_{\D_0}(s) = \frac{1}{(2k-1)!} \sum_{i=1}^{k-1} \a_i \sum_{\s=k-1}^{\infty}
\frac{2 \s + 1}{[\s (\s+1)]^{s-i}} \ ,
\label{zsca}
\ee
in terms of the coefficients $\a_i$ defined by 
\be
\frac{(\s+k-1)!}{(\s-k+1)!} = 
\prod_{i=0}^{k-2} \left[ \s (\s+1) -i(i+1) \right] =
\sum_{i=1}^{k-1} \a_i [\s (\s+1)]^i \ ,\qquad (k\ge 2)\ ,
\label{a3}
\ee
where $\s = l+k-1$. We use the analytic continuation (\ref{weized}) for all
the terms in (\ref{zsca}), which reduce to finite sums
involving the Bernoulli numbers $B_n$ in the limit $s=0$:
\be
\z_{\D_0}(0) = - \frac{1}{(2k-1)!} \sum_{i=1}^{k-1} \a_i\ \sum_{m=0}^{i+1}
 \frac{i!}{m!(i-m+1)!}B_{2i+2-m} \ .
\label{a6}
\ee
This expression can be cast into  the following compact form:
\be
\z_{\D_0}(0) = - \frac{1}{(2k-1)!} \int_{0}^{B(1+B)} dt \prod_{i=0}^{k-2}
\left[t-i(i+1) \right] \ , \qquad\quad (d=2k \ge4),
\label{zd0}
\ee
where it is understood that 
$B^n$ should be replaced with $B_n$ after evaluation of the integral. 
The values of $\z_{\D_0}$ for the lowest dimensions
$d=\left\{4,6,\dots,14 \right\}$ are:
\be
\z_{\D_0}(0)= \left\{- \frac{1}{90}\ ,\ \frac{1}{756}\ ,\ -
\frac{23}{113400}\ ,\ \frac{263}{7484400}\ ,\ -
\frac{133787}{20432412000}\ ,\ \frac{157009}{122594472000} \right\}\ .
\label{sample0}\ee

Finally, the trace anomaly of the scalar
field is obtained by replacing (\ref{zd0}) into (\ref{thedef})
(the result is in agreement with Ref.\cite{ct1}).
We postpone the discussion of these numbers to  Section $4$
and move to the analogous calculation for the Dirac field.


The spinor Laplacian is obtained from the square of the Dirac
operator in the $g_{\m \n}$ background,
\be
\D_{1/2} = - \fey{\nabla}^{ \ 2} = -\nabla^2 + \frac{\cal R}{4} \ , 
\label{a81}
\ee
and it is conformally covariant \cite{bd};
its eigenvalues and degeneracies on the sphere are \cite{cw}:
\ba
\l_l &=& (l+k)^2 \quad , \ \hspace{3cm} l = 0, 1, 2, ... \nonumber \\
\d_l &=& 2^{k+1} \frac{(2k+l-1)!}{l!(2k-1)!}   \quad .
\label{a9}
\ea
The analytic continuation of the corresponding zeta-function is again
obtained by splitting it into simpler functions ($k\ge 1$):
\be
\z_{\D_{1/2}}(s) = \frac{2^{k+1}}{(2k-1)!} \sum_{l=0}^{\infty} 
\frac{(l+2k-1)!}{l!} \frac{1}{(l+k)^{2s}} =
\frac{2^{k+1}}{(2k-1)!} \sum_{i=0}^{k-1}\ \b_i \ \z(2s-2i-1)\ ,
\label{a91}
\ee
where the coefficients $\b_i$ are defined by
\be
\frac{(\s+k-1)!}{(\s-k)!} = 
\s \prod_{i=1}^{k-1} \left(\s^2 -i^2 \right) =
\sum_{i=0}^{k-1}\ \b_i\ \s^{2i+1} \quad ,
\label{a93}
\ee
with $\s=l+k$. The limit $s=0$ can be taken into Eq.(\ref{a93})
and the result can be written again in compact form:
\be
\z_{\D_{1/2}}(0) = -
\frac{2^k}{(2k-1)!}\ \int_{0}^{B^2} dt \prod_{i=1}^{k-1}
(t-i^2) \ , \qquad\qquad \left( B^n \to B_n\right) \ 
\label{a10}
\ee
(for $k=1$ the integrand is one).
The first few values for $d=\left\{4,6,\dots,14\right\}$ are:
\be
-\ \z_{\D_{1/2}}(0) = \left\{- \frac{11}{90}\ ,\ \frac{191}{3780} \ ,
 - \frac{2497}{113400}\ ,\ \frac{14797}{1496880}\ ,  
- \frac{92427157}{20432412000}\ ,\ 
\frac{36740617}{17513496000} \right\} .
\label{sample1}\ee



\section{Trace anomaly of the antisymmetric tensor field}

Another conformal invariant free theory in 
$2k$ dimensions is given by the antisymmetric $(k-1)$-form, called 
antisymmetric tensor field \cite{ghost}, 
which generalizes the four-dimensional vector field.
Antisymmetric tensor fields commonly appear in supergravity and in 
superstring theory; 
a well known example \cite{m5} is the low-energy world-volume theory 
of the $M5$-brane described by a $d=6$, ${\cal N}=(2,0)$ tensor multiplet that 
contains one antisymmetric tensor $B_{\m \n}$.
 
In general, one could consider any $p$-form $A$, with $p=1,\dots,k-1$, 
and write its action in terms of the field strength $F=dA$ as follows
(using the notations of Ref.\cite{ct2}):
\be
S = - \frac{1}{2p!} \int F \wedge\ \ast F \quad ;
\label{c1}
\ee
however, this is conformal invariant for $p=k-1$ only.
For general $p$, this action is invariant under the gauge
transformation  $\ A \to A + d\m\ $, with 
$\ \m\ $ a $\ (p-1)$-form, thus we should introduce gauge-fixing terms 
and ghost fields. 
The ghost action is also gauge invariant, so we should consider
ghosts of ghosts: the resulting tower of ghost fields contains all
the forms of lower degree, till the $0$-forms which have no further gauge 
invariance.
This quantization problem is well understood \cite{ghost}\cite{ct2} 
and the resulting quantum action takes the form: 
\be
S = -\frac{1}{2} \sum_{i=0}^{p} \frac{1}{(p-i)!} \int A_{p-i} \wedge
\ \ast\D_{p-i}^{i+1}\ A_{p-i} \quad ,
\label{c2}
\ee
where the $(p-i)$-form $A_{p-i}$ for $i$ even (odd)
is a bosonic (fermionic) field and $\D_{p-i}$ is the Hodge-de Rham 
operator $\D = d \d + \d d \ $ expressed in terms of the 
exterior derivative $d$ and the coderivative $\d$. 
It follows that the partition function for the antisymmetric tensor
theory ($p=k-1$) in $d=2k$ is given by the following product 
of determinants:
\be 
Z_{AT} = \frac{1}{\det^{1/2}(\D_{k-1})} \cdot 
\frac{\det(\D_{k-2})}{\det^{3/2}(\D_{k-3})} \cdots
\left( \frac{\det^{(k-1)/2}(\D_1)}{\det^{k/2}(\D_0)} \right)^{(-1)^{k-1}} 
\quad .
\label{b0}
\ee

On the geometry of the sphere, the eigenvalues and degeneracies for exact 
$(p+1)$-forms and coexact $p$-forms coincide \cite{ct2}, so we can 
simplify the ratios in the previous expression 
by restricting the determinants to the spectra of coexact forms:
\be
Z_{AT} =  \left[\ 
\frac{1}{\det(\D^{ce}_{k-1})}\cdot
\frac{\det(\D^{ce}_{k-2})}{\det(\D^{ce}_{k-3})}\cdots
\left( \frac{\det(\D^{ce}_1)}{\det(\D^{ce}_0)}\ V_{2k} \right)^{(-1)^{k-1}}
\ \right]^{1/2} \quad ,
\label{b1}
\ee
where the superscript in $\D^{ce}$ means the restriction to coexact forms.
Note the volume factor in the last term that arises from
the regularization of the zero 
mode of the $0$-form, which does not cancel out in the last 
ratio of Eq.(\ref{b0}).

The eigenvalues and degeneracies of the Hodge-de Rham operator on 
coexact $p$-forms in $S^{2k}$ are given by \cite{ct2}:
\ba
\l_{p,l} &=& (l+p)(l+2k-p-1) \quad , \hspace{4cm} l = 1, 2, ... \nonumber \\
\d_{p,l} &=& \frac{(l+2k-1)!}{p!(2k-p-1)!(l-1)!} 
             \frac{2l+2k-1}{(l+p)(l+2k-p-1)}  \  .
\label{b2}\ea
The calculation of the trace anomaly thus involves the following 
alternate sum of zeta-functions:
\be
{d \over d\a} \log Z_{AT}\equiv
\z_{AT}(0) = \sum_{p=0}^{k-1} (-1)^{k-p-1}\z_{\D_p^{ce}}(0) \  
+ \ (-1)^{k-1}k \ .
\label{b3}
\ee
Each zeta-function can be analytically continued with the help
of Eq. (\ref{weized}); after some simplifications, the result reads: 
\be
\z_{AT}(0) =  \sum_{q=1}^{k}\frac{(-1)^{q}}{(k-q)!(k+q-1)!}
\sum_{i=1}^{k}\a_{i}(q)\sum_{m=0}^i(2q-1)^m\frac{(i-1)!}{m!(i-m)!}
B_{2i-m} \ ,
\label{b4}
\ee
where the coefficients $\a_i(q)$, with $q=k-p$, are defined by
the expression:
\ba
\frac{(\s +k)!}{(\s -k)!} & = & 
\prod_{i=0}^{k-1}\left[(\s +q)(\s -q+1) -i(i+1)-q(1-q)\right] \nl
& = & \sum_{i=1}^k\ \a_i(q)\ \left[ (\s +q)(\s -q+1) \right]^i \ .
\label{b5}
\ea
The result (\ref{b4}) can again be written in the compact form 
(using $ B^n \to B_n $):
\be
\z_{AT}(0) = 
\sum_{q=1}^{k}\frac{(-1)^{q}}{(k-q)!(k+q-1)!}
\int_{0}^{B(B+2q-1)}\ \frac{dt}{t} \ 
\prod_{i=0}^{k-1}\left[ t-i(i+1)-q(1-q) \right] \ .
\label{b6}
\ee
The first few values of this quantity 
for $d=\left\{4,6,\dots,12\right\}$ are:
\be
\z_{AT}(0) = \left\{ - \frac{62}{90}\ ,\ \frac{221}{210} \ ,\ 
- \frac{8051}{5670}\ ,\ \frac{1339661}{748440}\ ,\ 
 - \frac{525793111}{243243000}\ ,\ 
\frac{3698905481}{1459458000} \right\}\ .
\label{sampleAT}\ee
For $k=2$ this matches the well-known trace
anomaly of the vector field \cite{bd}, and for $k=3$ it checks
the recent result for the two-form gauge field in Ref.\cite{bft}.


\section{Anomaly versus number of degrees of freedom}

The trace anomaly in $2k$ dimensions contains the Euler density $G$
and a number of conformal covariant polynomials which involve 
the Weyl tensor $W$ and its derivatives, and have dimension $2k$ \cite{ds}:
\be
\la \Theta(x)\ra = \l \left( a\ G(x)\ -
\ c\ W(x)\ \D^{k-2}\ W(x)\ +\ O(W^3)+\dots 
\right) \quad .  
\label{anomdef}\ee
In this Equation, we called $a$ the coefficient of the Euler
density $G$ and $c$ that of the term quadratic in the Weyl tensor,
with $\D$ a conformal covariant Laplacian, and we left unspecified
the other conformal covariant polynomials.
The general form (\ref{anomdef}) can be integrated on conformally flat
manifolds ${\cal M}$ to yield:
\be
{d \over d \a} \log Z\left[{\cal M}\right] =
-{\l \over V_{2k-1}}\ a\ \chi =
\z_{\D} (0)\ {\chi \over 2}  \ ,
\label{chieq}\ee
with $\chi=\int G$ the Euler characteristic.
In the right part of this Equation, we also matched 
the general expression with the previous calculation on the sphere, 
for which $\chi\left( S^{2k} \right)=2$. 
The normalization coefficient $\l$ remains to be chosen for 
the complete definition of $a$.

In the following, we shall argue that there exist an absolute, 
i.e. dimension and scale
independent  normalization for $a$ (call it $a=\hat{a}$),
owing to the topological invariance of the Euler characteristic,
and a relative, $d$-dependent normalization suitable for using 
$a$ as a counting function.

Let us discuss the first definition. The Euler characteristic 
$\chi\left({\cal M} \right)$ has a natural normalization because it
takes integer values:
it counts the number $b_p$ of non-trivial zero modes of the Hodge-De Rham
operator $\D_p$ on the manifold ${\cal M}$, more precisely 
their alternating sum $b_0-b_1+b_2-\cdots+b_{2k}$ .
On the other side of Eq.(\ref{chieq}), the scale derivative of
$\log Z$ is also a number free from ambiguities, being 
the finite part after regularization of the ``number of positive
modes'' of a given Laplacian on ${\cal M}$. 
The ratio of these two numbers is
an interesting dimension and scale independent 
quantity; thus, we identify it with $a=\hat{a}\equiv \z _{\D}(0)$,
upon choosing the normalization $\l = - V_{2k-1}/2 $.

This discussion suggests a similarity between the topological part of 
the trace anomaly and the chiral anomaly, which counts the number 
of certain zero modes on ${\cal M}$, as a consequence of the index
theorem for the Dirac operator \cite{lagw}.
We have seen that the Laplacians of the free fields in Sections $2$
and $3$ do not have zero modes on the sphere, 
but nevertheless their trace anomaly is proportional to $\chi$
which counts zero modes. 
Suppose instead that  $\D$ had $n$ zero modes: they would be
regularized by an infra-red cut-off $\r$ and would contribute to
the partition function a factor $Z\sim \r^n$, leading nicely 
to $\hat{a}=n$.
For the actual $\D$ without zero modes, we can think that the
ultra-violet regularization of the non-zero modes
produces ``effective'' zero modes.
In conclusion, $\hat{a}=\z_{\D}(0)$ can be interpreted as 
counting effective zero modes of the Laplacian of the theory.

\begin{table}[t]
\begin{center}
\[
\begin{array}{|c||c|c|c|c|c|c|c|c|}
\hline 
d      & 4    & 6    & 8    & 10    & 12    & 14   &       & 2k   \\ 
\hline\hline
a(S)   & 1    & 1    &   1  &   1   &   1   &   1  &       &      \\ 
\hline
a(F) & 11   & \frac{191}{5} &\frac{2497}{23}    &\frac{73985}{263}   
 &\frac{92427157}{133787}     & \frac{257184319}{157009}   & \cdots &  \\ 
\hline
a(AT) & 62   &\frac{3978}{5} &\frac{161020}{23}  & \frac{13396610}{263}   
&\frac{44166621324}{133787}   &\frac{310708060404}{157009} & \cdots &  \\ 
\hline\hline
r(S)   & 1    & 1    &   1  &   1   &   1   &   1  &       & 1         \\ 
\hline
r(F) & 2.75 & 4.77 & 6.79 &   8.79&  10.79& 12.80&\cdots & \simeq 2k\ \ \ \\ 
\hline
r(AT) & 31  & 132.6 & 350.0& 727.7 & 1310. &2142. &\cdots & \simeq (2k)^3 \\ 
\hline
\end{array}
\]
\end{center}
\caption{Counting function $a$ and the corresponding weight per field 
component $r$ in various dimensions $d$, with asymptotic
behaviours for $d\to\infty$.}
\label{atab}
\end{table}

We now discuss the dependence of this quantity on the dimension and the spin.
Equation (\ref{sample0}) shows some values of $\hat{a}$ for the scalar theory; 
one sees that their sign alternates with $k$ and that they
decrease (exponentially) with the dimension.
Actually, the large $d$ limit is equivalent to the semiclassical limit
for the scalar theory, and the anomaly should go to zero.
The semiclassical limit can be easily understood
by defining the theory on a space-time lattice:
the field variables are located on the lattice
sites, and the discretized Laplacian of the field at one point 
$\D \f (x_o)$
is the sum of all the $2d$ field variables on the nearest neighbour
sites. As the value of $d$ increases, 
the action approaches its mean-field approximation.

By the same argument we expect that the values of $\hat{a}$ 
for the fermion and antisymmetric tensor fields should
also vanish for large $d$. This is indeed the case:
we should take the values in Eqs.(\ref{sample1},\ref{sampleAT}) 
and divide them for the corresponding number of on-shell field components 
(number of independent polarizations), which are \cite{ghost}:
\be
n(F)= 2^k\ ,\qquad\qquad n(AT)={ (2k-2)! \over
\left[ (k-1)! \right]^2 }\ .
\ee
In conclusion, the trace anomaly $\hat{a}$ is a measure of
effective zero modes and decreases for large $d$ in absolute
units as expected.

\bigskip

\begin{table}[t]
\begin{center}
\[
\begin{array}{|c||c|c|c|c|c|c|c|c|}
\hline 
d     & 4    & 6     & 8    & 10    & 12    & 14    &       & 2k     \\ 
\hline\hline
c(S)  & 1    & 1     &   1  &   1   &   1   &   1   &       &        \\ 
\hline
c(F)& 6    & 20    & 56   & 144   & 352   & 832   &\cdots &        \\ 
\hline
c(AT)& 12   & 90    & 560  & 3150  & 16632 & 84084 &\cdots &        \\ 
\hline\hline
r'(S) & 1    & 1     &   1  &   1   &   1   &   1   &       &  1     \\ 
\hline
r'(F)&\frac{3}{2} & \frac{5}{2} &\frac{7}{2} &\frac{9}{2} 
       &\frac{11}{2}& \frac{13}{2}&  \cdots    &  \simeq k \ \ \       \\ 
\hline
r'(AT)& 6   & 15    &   28 &   45  &   66  &   91  &\cdots&\simeq 2k^2 \\ 
\hline
\end{array}
\]
\end{center}
\caption{Counting function $c$ and the corresponding weight per field 
component $r'$ in various dimensions $d$, with asymptotic
behaviours for $d\to\infty$.}
\label{ctab}
\end{table}

We now analyse the coefficient $a$ as a measure of degrees
of freedom in field theory; this interpretation would be
implied by the eventual proof of the $c$-theorem.
In this case, we should normalize $a(S)=1$ for the scalar
field in any dimension, by choosing $\l$ in (\ref{chieq}) accordingly.
The values of $a$ in this normalization are reported in 
the first three lines of Table(\ref{atab}); 
the last three lines instead contain the relative weights per field
component:
\be
r(\s) ={a(\s) \over n(\s)}\ , \qquad \qquad \s= S,\ F,\ AT\ .
\label{rdef}\ee
The last column in Table (\ref{atab}) gives their asymptotic behaviour
for large dimension, which were determined numerically from 
Eqs.(\ref{zd0},\ref{a10},\ref{b6}) (more precisely, 
$r(AT) \simeq 0.943 \ (2k)^3$).
We find that these ratios do not go to one, as we would naively
expect in the semiclassical limit, but grow, in particular for the
antisymmetric tensor: already at $d=4$ the vector field
has a rather large weight.
Therefore, this ``quantum'' measure of degrees of freedom is always
different from the classical number of field components -- a fact that
is rather counter-intuitive.

Apart from this physical interpretation, the behaviour of $a$ is not
surprising because it also occurs for other terms in the trace
and gravitational anomalies.
This means that the spinful theories approach the 
semiclassical limit more slowly than the scalar theory, or that they 
interact more strongly with the background metric.
Let us first discuss the other trace anomaly coefficient $c$ 
which was introduced
in Eq. (\ref{anomdef}); this number also gives the normalization 
of the stress tensor correlator in flat space \cite{cc} and
has been already computed in the Refs. \cite{cc}\cite{c-at}
for the free theories considered here.
In the same normalization $c(S)=1$, the results are: 
\be
c(F)= 2^{k-1}(2k-1) \ ,
\hspace{1cm} c(AT)= \frac{(2k)!}{2[(k-1)!]^2}\ ,\qquad (d=2k)\ .
\label{d1}
\ee
In Table (\ref{ctab}) we display the first few values of $c$
as well as the corresponding weights per field component
$r'(\s)=c(\s)/n(\s)$. As we anticipated,
these ratios grow with the dimension, although with milder
asymptotic behaviours.

The same enhancement is observed in the chiral anomaly of
pure gravitational origin, which occurs for dimensions $d=2+4n$.
The comprehensive study of Ref.\cite{lagw} considers the following
chiral fields: the Weyl fermion (one half
of the Dirac fermion), the gravitino
and the self-dual antisymmetric tensor (one half of that considered
here). 
The complete expression of this chiral anomaly contains several 
independent terms, made by the different traces of the product of 
$(2+2n)$ Riemann two-forms, and their coefficients
span a wide range of values \cite{lagw}. 
The larger coefficient is found
for the genuine $d=2+4n$ term, which is the $(n+1)$-th Pontryagin class
$p_{n+1}({\cal M}) = {\rm Tr} (R^{2n+2}) +\cdots$.

We have computed the corresponding weights per field component for 
the three field types considered in Ref. \cite{lagw} and for
$d=6,10,14$: these weights again grow with the dimension,
the faster the higher spin value; moreover, 
the relative growth between the antisymmetric tensor
and the fermion is much larger than the order $O(d^2)$
found for the coefficient $a$.
The origin of this enhancement is not immediately apparent
in the diagrammatic calculation of Ref.\cite{lagw},
because the coupling of spin to gravity does not have a standard
group-theoretic form valid for all spin values. 
Nevertheless, the number of external legs in the anomalous loops 
increase  with the dimension and the
algebraic complexity enhances the spin dependence.


\section{Discussion}

Although puzzling, the measure of degrees of freedom given by
$a$ could be consistent with the renormalization
group flow. Here we can check a couple of 
standard renormalization-group patterns and discuss a $c$-theorem conjecture
\cite{c-a} which is consistent with the observed enhancement.

A well-known flow is provided by a (generalized) gauge theory which
has massive gauge fields in the infrared, together with massless scalars 
due to spontaneous symmetry breaking of a global symmetry;
our results imply a very large value for $a_{UV}$, which better fulfills
the $c$-theorem inequality $a_{UV} > a_{IR}$.
Another example is the
Higgs phenomenon, in which scalar degrees of freedom become
extra components for the massive antisymmetric tensor.
For large $d$, these extra components contribute to $a$ far less than
the original components of the massless field:
anyhow, no contradiction arises in going back to the UV limit,
because the massless limit of the massive gauge theory is 
different from the massless theory.

In the Reference \cite{c-a}, it has been conjectured 
that the canonical form of the two-dimensional $c$-theorem could extend
to the class of higher dimensional theories characterized by $c=\nu\ a$, with
$\nu(d)$ a given constant.
The field theories described by the AdS/CFT
correspondence belong to this class 
\cite{hs}\cite{agpz}; for instance, the vector multiplet of 
${\cal N}=4$ supersymmetry in four dimensions.
In our notations $a(S)=c(S)=1$, the
equation $c=\nu a$ of Ref. \cite{c-a} reads:
\be
c\ - (2k+1)\ \frac{\G (2k)}{\G (k)^2}\ 
\left\vert \z_{\D_0}(0) \right\vert\ a =0 \ , \qquad \qquad (d=2k)  \ .
\label{d2}
\ee
Since the scalar, fermion and antisymmetric tensor fields contribute
differently to $a$ and $c$, and with definite sign, there
is no {\it a-priori} guarantee that this equation admits solutions 
in higher dimensions.

We can easily obtain the explicit form of this equation in each dimension 
by using the results in the two Tables; for $d=4,6,8,10$, we find:
\be
\begin{array}{rrrrrrr}
d = 4\ :    &&& 2\ N_S \  +& 7\ N_F   \  -& 26\ N_{AT}    \   =& 0 \ ; \\
d = 6\ :  &&& 13\ N_S  \ +& 169\ N_F \  -& 2358\ N_{AT}    \ =& 0 \ ;  \\
d = 8\ :  &&& 67\ N_S  \ +& 2543\ N_F \ -& 110620\ N_{AT}  \ =& 0 \ ; \\
d = 10 : &&& 817\ N_S  \ +& 81535\ N_F \ -& 9994610\ N_{AT}  \ =& 0 \ ,
\end{array}
\label{int}
\ee
where $(N_S,N_F,N_{AT})$ are the field multiplicities.
We see that the dominance of $a(AT)$ 
yields a crucial minus sign in these equations,
which ensures a solution in any dimension.
Therefore, we have found that the behaviour of $a$ 
makes the $c$-theorem conjecture possible in any dimension. 

The integer equations ( \ref{int}) has 
two independent solutions $(N_S,N_F,N_{AT})$,
which can be thought of as being vectors generating a 
two-dimensional sub-lattice of a three-dimensional integer lattice.
Some sample solutions, with minimal values for $N_{AT}$,
are the following:
\be
\begin{array}{llll}
d= 4\ : & & (6,2,1), & (13,0,1); \nonumber \\
d= 6\ : & & (161,169,13), & (265,161,13); \nonumber \\
d= 8\ : & & (835,65,2),   & (2524,64,3); \nonumber \\
d= 10\ : & & (11950,248,3), & (47095, 141,5)\ . 
\end{array}
\label{solint}\ee

\vskip 24pt
\begin{flushleft}
{ \bf Acknowledgments}
\end{flushleft}
We thank D. Anselmi, A. Coste, R. Guida, N. Magnoli and A. Schwimmer
for interesting discussions on the $c$-theorem and the trace anomaly.
This work is supported in part by the European Community 
Network grant FMRX-CT96-0012.
 
\def\NPB#1#2#3{{\it Nucl.~Phys.} {\bf{B#1}} (#2) #3}
\def\CMP#1#2#3{{\it Commun.~Math.~Phys.} {\bf{#1}} (#2) #3}
\def\CQG#1#2#3{{\it Class.~Quantum~Grav.} {\bf{#1}} (#2) #3}
\def\PLB#1#2#3{{\it Phys.~Lett.} {\bf{B#1}} (#2) #3}
\def\PRD#1#2#3{{\it Phys.~Rev.} {\bf{D#1}} (#2) #3}
\def\PRL#1#2#3{{\it Phys.~Rev.~Lett.} {\bf{#1}} (#2) #3}
\def\ZPC#1#2#3{{\it Z.~Phys.} {\bf C#1} (#2) #3}
\def\PTP#1#2#3{{\it Prog.~Theor.~Phys.} {\bf#1}  (#2) #3}
\def\MPLA#1#2#3{{\it Mod.~Phys.~Lett.} {\bf#1} (#2) #3}
\def\PR#1#2#3{{\it Phys.~Rep.} {\bf#1} (#2) #3}
\def\AP#1#2#3{{\it Ann.~Phys.} {\bf#1} (#2) #3}
\def\RMP#1#2#3{{\it Rev.~Mod.~Phys.} {\bf#1} (#2) #3}
\def\HPA#1#2#3{{\it Helv.~Phys.~Acta} {\bf#1} (#2) #3}
\def\JETPL#1#2#3{{\it JETP~Lett.} {\bf#1} (#2) #3}
\def\JHEP#1#2#3{{\it JHEP} {\bf#1} (#2) #3}
\def\TH#1{{\tt hep-th/#1}}

\end{document}